\documentclass[fleqn,twoside]{article}
\usepackage{espcrc2}
\usepackage{graphicx}
\usepackage{amsfonts}

\title{Lattice regularized chiral perturbation theory}

\author{Bu\=gra Borasoy\address{Physik Department, Technische Universit\"at
        M\"unchen, D-85747 Garching, Germany},
        Randy Lewis\address{Department of Physics, University of Regina,
                            Regina, SK, S4S 0A2, Canada},
        and Pierre-Philippe A. Ouimet$^{\rm b}$
        }
\begin{document}

\begin{abstract}
Chiral perturbation theory can be defined and
regularized on a spacetime lattice.  A few motivations are discussed here,
and an explicit lattice Lagrangian is reviewed.
A particular aspect of the connection between lattice chiral perturbation
theory and
lattice QCD is explored through a study of the Wess-Zumino-Witten term.
\end{abstract}

\maketitle

\section{MOTIVATIONS}

The spacetime lattice is a regularization technique
for quantum field theories.
Interest in the lattice regularization of chiral perturbation theory (ChPT)
has arisen in a number of different contexts.

\subsection{Discretization effects for lattice QCD}

To account for the effects of nonzero lattice spacing, $a$, in lattice
QCD simulations, work has been done to include explicit $a$-dependences into
the effective field theory, ChPT.  Since $1/a$ in lattice QCD simulations is
typically larger than the chiral scale $\Lambda_\chi \sim 1$ GeV,
it is common to include these $a$-dependences
by adding extra ``irrelevant'' operators to ChPT
\cite{Baer,Aoki1,Beane}.  Calculations are then performed in the continuum,
using dimensional regularization or some other continuum regularization method,
and results contain both $a$-dependent terms and $\mu$-dependent
terms where $\mu$ is the regularization scale.  As usual, physical observables
will be independent of $\mu$ in a regime where ChPT is applicable
and has acceptable convergence properties.

Alternatively, one might prefer to define lattice ChPT as an effective theory
that exists directly in the same
discrete spacetime where lattice QCD resides\cite{Levi,Shushpanov,BLO}.
With this
approach the lattice itself regulates the theory, playing the roles that were
assigned to
both $\mu$ and $a$ in the continuum method discussed above.  This type of
lattice ChPT has
no divergences for nonzero $a$ and it has no need for a continuum regulator.
With small lattice spacings, i.e.~$1/a\gg\Lambda_\chi$,
physical observables will be essentially independent of the regularization
scheme and scale, so the results will reproduce those
obtainable from the continuum
theory of the previous paragraph (up to negligible differences that vanish
exactly as $a\to0$).

On coarse lattices, perhaps $1/a\sim\Lambda_\chi$, it is important to
determine whether the scheme dependence really is acceptably negligible before
relying on a particular regularization scheme.
Explicit lattice
ChPT studies of physical observables over a range of lattice spacings
can help to quantify how small the lattice spacing must be to ensure that
this type of scheme dependence is indeed negligible.

\subsection{Convergence and power divergences}

Lattice regularized ChPT is also of interest independently of its
connection to lattice QCD.  There has recently been renewed interest in
cut-off type regularization schemes for ChPT\cite{UMass,Adelaide,BHM}, and
in this context lattice regularized
ChPT might be viewed as yet another way to implement a cut-off.

Ref.~\cite{UMass} contains some studies of the convergence of
ChPT by using a particular momentum cut-off scheme, referred to as
long-distance regularization, and studying the cut-off dependences
order by order in the chiral expansion.  Although each
observable needs to be essentially independent of regularization scheme,
the convergence of ChPT is typically defined in a scheme-dependent fashion by
comparing the relative sizes of the contributions to some observable at each
chiral order.
Like most regularization schemes (but unlike dimensional regularization)
cut-off schemes have power divergent loop integrals, and variation
of the cut-off can significantly change the size of loop contributions relative
to counterterms.
Such studies provide insight into the convergence of ChPT
and into the effect of having to truncate ChPT at some fixed chiral order.

The Adelaide group has discussed scheme dependence and convergence
for a collection of ``finite range regulators''\cite{Adelaide}, similar to
the long-distance regularization scheme mentioned above, and they have made
extensive studies of their method in the context of
practical extrapolations for lattice QCD data.

Lattice regularized ChPT is also a method for invoking a cut-off, with its
own specific features.
A potential disadvantage of the lattice method
is that the continuous rotational symmetry is reduced to a hypercubic
rotational symmetry, though this only becomes significant at the very
short distance scale set by the lattice spacing.  An advantage of the lattice
method is that the lattice
spacing appears directly in the Lagrangian, in contrast to schemes where the
cut-off is inserted by hand into each loop integration.  Therefore every
calculation with the lattice method automatically preserves all of the
Lagrangian's symmetries (like chiral symmetry and gauge invariance) whereas
the users of other cut-off methods must be careful to ensure that
implementation of the cut-off preserves the desired symmetries during the
calculation of each loop integral.

\subsection{Nonperturbative issues}

A particularly valuable feature of lattice regularization is that it does
not rely on perturbation theory.  Lattice QCD takes advantage of this, and
a multi-nucleon effective field theory might also find
lattice regularization to be a useful tool.\cite{Chandrasekharan}
In the present work, we will restrict ourselves to ChPT in the presence of
at most a single baryon.

\section{MESON SECTOR}

Chiral perturbation theory is an expansion in inverse powers of
$\Lambda_\chi \sim 4\pi{F_\pi} \sim 4\pi{F_K} \sim 1$ GeV.
In Euclidean spacetime, the familiar chiral Lagrangian\cite{Gasser}
for pseudoscalar mesons is
\begin{eqnarray}
{\cal L}_{\rm M} &\!=\!& {\cal L}_{\rm M}^{(2)} + {\cal L}_{\rm M}^{(4)}
                   + {\cal L}_{\rm M}^{(6)} + \ldots, \label{GLLagrangian1} \\
{\cal L}_{\rm M}^{(2)}  &\!=\!& \frac{F^2}{4}{\rm Tr}(\sum_\mu
   \nabla_\mu{U}^\dagger\nabla_\mu{U}-\chi^\dagger{U}-\chi{U}^\dagger), \\
{\cal L}_{\rm M}^{(4)} &\!=\!& -L_1\left(\sum_\mu{\rm Tr}(
   \nabla_\mu{U}^\dagger\nabla_\mu{U})\right)^2
\nonumber \\
&& - L_2\sum_{\mu,\nu}{\rm Tr}(\nabla_\mu{U}^\dagger\nabla_\nu{U})
   {\rm Tr}(\nabla_\mu{U}^\dagger\nabla_\nu{U}) \nonumber \\
&& - L_3\sum_{\mu,\nu}{\rm Tr}(\nabla_\mu{U}^\dagger\nabla_\mu{U}
   \nabla_\nu{U}^\dagger\nabla_\nu{U})
\nonumber \\
&& + L_4\sum_\mu{\rm Tr}(\nabla_\mu{U}^\dagger\nabla_\mu{U}\}
   {\rm Tr}\{\chi^\dagger{U}+\chi{U}^\dagger) \nonumber \\
&& + L_5\sum_\mu{\rm Tr}(\nabla_\mu{U}^\dagger\nabla_\mu{U}
   (\chi^\dagger{U}+U^\dagger\chi))
\nonumber \\
&& - L_6\left({\rm Tr}(\chi^\dagger{U}+\chi{U}^\dagger)\right)^2 \nonumber \\
&& - L_7\left({\rm Tr}(\chi^\dagger{U}-\chi{U}^\dagger)\right)^2
\nonumber \\
&& - L_8{\rm Tr}(\chi^\dagger{U}\chi^\dagger{U}+\chi{U}^\dagger\chi{U}^\dagger
     ) \nonumber \\
&& + iL_9\sum_{\mu,\nu}{\rm Tr}(F^R_{\mu\nu}\nabla_\mu{U}\nabla_\nu{U}^\dagger
     \nonumber \\
&&   +F^L_{\mu\nu}\nabla_\mu{U}^\dagger\nabla_\nu{U})
\nonumber \\
&& - L_{10}\sum_{\mu,\nu}{\rm Tr}(U^\dagger{F}^R_{\mu\nu}UF^L_{\mu\nu}),
\label{GLLagrangian2}
\end{eqnarray}
where the fields are
\begin{eqnarray}
U(x) &=& \exp\left[\frac{-i\lambda^a\pi^a(x)}{F}\right], \\
\chi &=& 2B\left(\begin{array}{ccc} m_u & 0 & 0 \\ 0 & m_d & 0 \\ 0 & 0 & m_s
         \end{array}\right) + \ldots, \\
L_\mu(x) &=& \exp[-ia\ell_\mu(x)], \\
R_\mu(x) &=& \exp[-iar_\mu(x)].
\end{eqnarray}
This Lagrangian is invariant under a local chiral transformation,
\begin{eqnarray}
U(x) & \to & g(x)U(x)h(x), \\
R_\mu(x) & \to & g(x)R_\mu(x)g^{\dagger}(x+a_\mu), \\
L_\mu(x) & \to & h^\dagger(x)L_\mu(x)h(x+a_\mu),
\end{eqnarray}
if the covariant derivatives are defined
appropriately.
To avoid extra (unphysical) states in the dispersion relation, 
a nearest-neighbour derivative will be used in ${\cal L}_{\rm M}^{(2)}$,
\begin{equation}
a\nabla_{\mu}^{(+)}U(x) = R_\mu(x)U(x+a_\mu)L_\mu^\dagger(x) - U(x).
\end{equation}
Use of this derivative at higher chiral orders would not preserve parity, so
a symmetric derivative is used everywhere except ${\cal L}_{\rm M}^{(2)}$,
\begin{eqnarray}
2a\nabla_{\mu}^{(\pm)}U(x)
   &\!\!\!\!=\!\!\!\!& R_\mu(x)U(x+a_\mu)L_\mu^\dagger(x) \\
   &\!\!\!\!-\!\!\!\!& R_\mu^\dagger(x-a_\mu)U(x-a_\mu)L_\mu(x-a_\mu).\nonumber
\end{eqnarray}

It should be emphasized that the lattice ChPT Lagrangian obtained by putting
these lattice derivatives into Eqs.~(\ref{GLLagrangian1}-\ref{GLLagrangian2})
is not the most general one that could be written on a hypercubic lattice;
it is merely one example of a ChPT Lagrangian that has the correct
continuum limit.  Similarly, there is an entire family of lattice QCD
Lagrangians that approach the unique continuum QCD as $a\to0$.
As discussed in section 1, extra irrelevant operators
that contain explicit powers of $a$ can be added to the ChPT Lagrangian of
Eqs.~(\ref{GLLagrangian1}-\ref{GLLagrangian2}) and
a particular choice for the set of associated coefficients would correspond to
a particular choice for the underlying lattice QCD Lagrangian.
Since our intent is mainly to study regularization issues,
we will arbitrarily set all of these extra coefficients to zero for simplicity.

Our full action contains the usual Lagrangian term plus a less familiar 
term which arises from the integration measure\cite{Rothe},
\begin{eqnarray}
S_{\rm M} &=& a^4\sum_x{\cal L}_{\rm M}(x) \nonumber \\
      &&  -\frac{1}{2}\sum_x{\rm Tr}\ln\left[\frac{2(1-\cos\Phi(x))}
                  {\Phi^2(x)}\right]
\end{eqnarray}
where
$\Phi(x) = \frac{2}{F}\sum_{a=1}^8t^a\pi^a(x)$
with $t^a_{bc}=-if_{abc}$.

For a sample calculation, consider the meson masses.
Neglecting isospin violation ($m_l \equiv m_u = m_d$),
the lowest order pion, kaon and eta two point functions are
\begin{equation}
\Gamma_{MM} =
-\left[x_M^2+\frac{4}{a^2}\sum_\mu\sin^2\left(\frac{aq_\mu}{2}\right)\right]
\end{equation}
where $M=\pi$, $K$ or $\eta$ and
\begin{eqnarray}
x_\pi &=& \sqrt{2Bm_l}, \\
x_K &=& \sqrt{B(m_l+m_s)}, \\
x_\eta &=& \sqrt{\frac{2}{3}B(m_l+2m_s)}.
\end{eqnarray}
The meson masses are therefore
\begin{equation}
m_M = \frac{2}{a}{\rm arcsinh}\left(\frac{ax_M}{2}\right).
\end{equation}
Notice the existence of a Gell-Mann--Okubo relation
\begin{eqnarray}
3{\rm sinh}^2\left(\frac{am_\eta}{2}\right)
= 4{\rm sinh}^2\left(\frac{am_K}{2}\right)
- {\rm sinh}^2\left(\frac{am_\pi}{2}\right) &&\hspace{-3cm}\nonumber \\ &&\hspace{-3cm}
\end{eqnarray}
which reproduces the conventional relation as $a \rightarrow 0$.
The next-to-leading order expressions include one-loop diagrams and
${\cal L}_{\rm M}^{(4)}$ tree-level pieces.
Loop momentum integration is from $-\pi/a$ to $+\pi/a$, representing the
complete range of momenta available in the lattice theory.
The results have the form
\begin{eqnarray}
\Gamma_{MM} &\!\!\!\!=\!\!\!\!& \frac{-1}{Z_M^{(+)}Z_M^{(\pm)}}\left[X_M^2+\frac{4Z_M^{(\pm)}}
              {a^2}\sum_\mu\sin^2\left(\frac{aq_\mu}{2}\right)\right.
               \nonumber \\
          &\!\!\!&  \left. + \left(\frac{1-Z_M^{(\pm)}}{a^2}\right)
              \sum_\mu\sin^2\left(aq_\mu\right)\right],
\end{eqnarray}
where $X_M$, $Z_M^{(+)}$ and $Z_M^{(\pm)}$
contain five Lagrangian parameters, $L_4$, $L_5$, $L_6$, $L_7$, $L_8$,
and a single integral,
\begin{equation}
W_4(\epsilon) \equiv \int_0^\infty{\rm d}x\,I_0^4(x)\exp\left[-x\left(4+
\frac{\epsilon}{2}\right)\right]
\end{equation}
where $I_0(x)$ is a Bessel function.
For example, the kaon mass is 
\begin{equation}
m_K = \frac{2}{a}{\rm arcsinh}\left(\frac{aX_K}{2}\right)
\end{equation}
where
\begin{eqnarray}
X_K^2 &\!\!\!=\!\!\!& x_K^2 - \frac{8}{F^2}x_K^2(x_\pi^2+2x_K^2)(L_4-2L_6)
      \nonumber \\
     &\!\!\!& - \frac{8}{F^2}x_K^4(L_5-2L_8) + \frac{7x_K^2}{24a^2F^2}
      \nonumber \\
     &\!\!\!& + \frac{x_K^2}{6a^2F^2}W_4(a^2x_\eta^2)
        - \frac{3x_\pi^2x_K^2}{64F^2}W_4(a^2x_\pi^2) \nonumber \\
     &\!\!\!& - \frac{3x_K^4}{32F^2}W_4(a^2x_K^2)
        - \frac{x_K^2x_\eta^2}{192F^2}W_4(a^2x_\eta^2).
\end{eqnarray}
Notice that the kaon mass vanishes in the chiral limit ($m_l = m_s = 0$),
consistent with the fact that the theory does indeed have exact chiral
symmetry even for $a\neq0$.

As $a\to0$, loop integrals diverge but the infinities can be absorbed into
renormalized parameters.  The meson masses and the scale dependences of
the counterterms are in analytic agreement with dimensional regularization.
For example, defining $\Delta L_i \equiv L_i^r(1/a_2)-L_i^r(1/a_1)$, one finds
\begin{eqnarray}
\Delta L_4-2\Delta L_6 &=&
-\frac{1}{36(4\pi)^2}\ln\left(\frac{a_2}{a_1}\right), \\
\Delta L_5-2\Delta L_8 &=&
\frac{1}{6(4\pi)^2}\ln\left(\frac{a_2}{a_1}\right), \\
3\Delta L_7+\Delta L_8 &=&
\frac{5}{48(4\pi)^2}\ln\left(\frac{a_2}{a_1}\right),
\end{eqnarray}
for sufficiently small lattice spacings $a_1$ and $a_2$.
These are precisely the same logarithms and numerical coefficients that
appear in dimensional regularization\cite{Gasser}.
Power divergences do arise in lattice regularized loop integrals, and they
always get absorbed into Lagrangian parameters during renormalization.

\section{HEAVY BARYON SECTOR}

Heavy baryon chiral perturbation theory (HBChPT) is an expansion in the inverse
baryon mass as well as the inverse chiral scale, $\Lambda_\chi$.
The heavy octet baryon field is
\begin{equation}
B_v(x) = \exp(im_{\rm HB}v\cdot{x})\frac{1}{2}(1+v\!\!\!/)B(x),
\end{equation}
where $m_{\rm HB}$ is chosen near the physical baryon mass.

The first few orders in the double expansion of Euclidean HBChPT are
\begin{eqnarray}
{\mathcal{L}}_{\rm MB} &\hspace{-3mm}=\hspace{-3mm}&
        {\mathcal{L}}_{\rm MB}^{(0)} + {\mathcal{L}}_{\rm MB}^{(1)}
      + {\mathcal{L}}_{\rm MB}^{(2)} + {\mathcal{L}}_{\rm MB}^{(3)}
      + \ldots, \\
{\mathcal{L}}_{\rm MB}^{(0)} &\hspace{-3mm}=\hspace{-3mm}&
 (m_0-m_{\rm HB}){\rm Tr}\left(\bar{B}_vB_v\right), \\
{\mathcal{L}}_{\rm MB}^{(1)} &\hspace{-3mm}=\hspace{-3mm}&
   \sum_\mu{\rm Tr}\bar{B}_v(v_\mu D_\mu B_v
   +{\mathcal{D}}S_\mu\{u_\mu,B_v\}
\nonumber \\ &\hspace{-1cm}&
+{\mathcal{F}}S_\mu[u_\mu,B_v]), \\
{\mathcal{L}}_{\rm MB}^{(2)} &\hspace{-3mm}=\hspace{-3mm}&
\frac{1}{2m_0}{\rm Tr}\bar{B}_v(v\cdot{D}v\cdot{D}-D^2)B_v
\nonumber \\ &\hspace{-1cm}&
-{\rm Tr}\bar{B}_v\left(b_{\mathcal{D}}\{\chi_+,B_v\}
+b_{\mathcal{F}}[\chi_+,B_v]\right) \nonumber \\
&\hspace{-10mm}&-b_0{\rm Tr}\left(\bar{B}_vB_v\right){\rm Tr}\left(\chi_+\right)
\nonumber \\ &\hspace{-1cm}&
 +\frac{i}{4m_0}\sum_{\mu,\nu}{\rm Tr}\bar{B}_v[S_\mu,S_\nu]\times
\nonumber \\ &\hspace{-1cm}&
(\mu_D\{\xi F_{\mu\nu}^L\xi^\dagger+\xi^\dagger F_{\mu\nu}^R\xi,B_v\}
\nonumber \\ &\hspace{-1cm}&
+\mu_F[\xi F_{\mu\nu}^L\xi^\dagger+\xi^\dagger F_{\mu\nu}^R\xi,B_v])
 +\ldots
\end{eqnarray}
where $\mathcal{D}$ and $\mathcal{F}$ are the axial couplings, and
$S_\mu=\frac{i}{2}\gamma_5\sum_\nu\sigma_{\mu\nu}v_\nu$. \\
The mesons are contained within
\begin{eqnarray}
U &=& \xi^2, \\
\chi_+ &=& \xi^\dagger\chi\xi^\dagger + \xi\chi^\dagger\xi, \\
u_\mu(x) &=& \frac{i}{2}\xi^\dagger(x)\nabla_\mu^{(\pm)}U(x)\xi^\dagger(x)
\nonumber \\
         &-& \frac{i}{2}\xi(x)\nabla_\mu^{(\pm)}U^\dagger(x)\xi(x).
\end{eqnarray}
It is convenient to choose the velocity parameter to have only a temporal
component, $v=(0,0,0,1)$, and to use a nearest-neighbour temporal derivative
$D_4$ but a next-nearest-neighbour spatial derivative $D_i$, since this
preserves parity without introducing unphysical states.
As was discussed for the meson Lagrangian,
the most general lattice HBChPT Lagrangian
will contain additional terms that have explicit powers of the lattice spacing.
The coefficients of these terms could be chosen by matching to a specific
version of lattice QCD, but here we set them all to zero and use the remaining
``minimal'' lattice ChPT that still has the unique continuum limit.

To incorporate the decuplet baryons as heavy fields, we use\cite{HHK}
\begin{eqnarray}
{\cal L}_{\rm MT}^{(1)} &=& -\sum_{\mu\nu}\sum_{ijk} \bigg(\bar{T}_{v\mu}^{ijk}
v_\nu D_\nu T_{v\mu}^{ijk} \nonumber \\
&&-{\cal H}\sum_l\bar{T}_{v\mu}^{ijk}S_\nu u_\nu^{kl}T_{v\mu}^{ijl}\bigg)
\nonumber \\ &&
 - \Delta\sum_\mu\sum_{ijk}\bar{T}_{v\mu}^{ijk}T_{v\mu}^{ijk}, \\
{\cal L}_{\rm MBT}^{(1)} &=& \frac{\cal C}{2}\sum_\mu\sum_{ijklm}
\epsilon_{ijk}(\bar{T}_{v\mu}^{klm}u_\mu^{mj}B_v^{li} \nonumber \\ &&
+\bar{B}_v^{il}u_\mu^{jm}T_\mu^{klm}).
\end{eqnarray}
The resulting spin-3/2 propagator is
\begin{equation}
\frac{-P^{3/2}_{\mu\nu}}{\Gamma_{TT}}
 = \frac{ia(\delta_{\mu\nu}-v_\mu v_\nu+\frac{4}{3}
         S_\mu S_\nu)}{\sin(ap_4)-i[a\Delta+2\sin^2(ap_4/2)]}.
\end{equation}

Calculations can now be done in a straightforward manner, just as was discussed
for the meson sector, and some specific calculations can be found in
Ref.~\cite{BLO}.
Loop integrals generally contain power divergences as $a\to0$
--- for an $O(p^3)$ calculation they can be cubic, quadratic and linear ---
but these can always be absorbed into Lagrangian parameters.
Gauge invariance is also automatically preserved, and an explicit calculation
is discussed in detail in Ref.~\cite{BLO}.

\begin{figure}
\includegraphics[height=55mm]{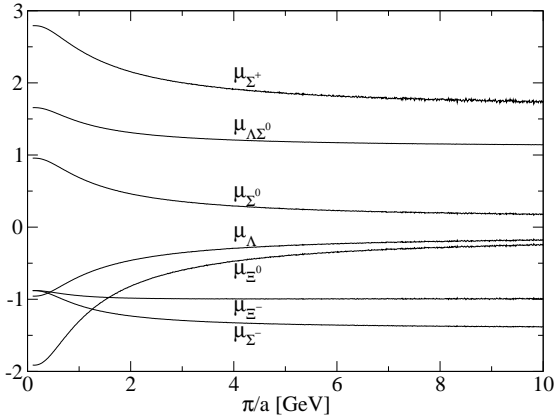}
\caption{Octet baryon magnetic moments from ChPT parameters that reproduce the
         experimental values for $\mu_p$ and $\mu_n$ at all lattice
         spacings.}\label{magmomfig}
\end{figure}
Figure \ref{magmomfig} shows the results of an $O(p^3)$ calculation of the
octet baryon magnetic moments as functions of the lattice spacing.  For this
illustrative example, parameters were chosen such that the proton and neutron
magnetic moments are equal to their experimental values at all lattice
spacings.  All of the remaining octet baryon
magnetic moments are plotted in the
figure, and Table \ref{magmomtab} shows the numerical
difference between results in the continuum and results at $a=0.1$ fm.

Two aspects of this illustrative example should be emphasized at this point.
First, it is well known that $O(p^4)$ corrections to the baryon magnetic
moments can be significant\cite{p4magmom}. Since our calculation is intended
to be a simple example of discretization effects, not a high precision
determination of the magnetic moments, we have not concerned ourselves
with the addition of $O(p^4)$ terms.
Secondly, the lattice spacing dependences computed here are, of course,
specific to the minimal version of lattice ChPT that we have employed, where
a collection of coefficients has been set to zero instead of being matched
to some specific lattice QCD Lagrangian.

\begin{table}
\caption{The numerical distinction between magnetic moments in the continuum
         and at $a=0.1$ fm, for a specific lattice ChPT
         Lagrangian.}\label{magmomtab}
\begin{tabular}{ccc}
$B$ & $\mu_B(a=0)$ & $\frac{\mu_B(a=0.1{\rm fm})}{\mu_B(a=0)}$ \\
\hline
$\Sigma^+$ & 1.64 & 1.11 \\
$\Sigma^0$ & 0.12 & 1.94 \\
$\Sigma^-$ & -1.40 & 0.97 \\
$\Xi^0$ & -0.14 & 2.58 \\
$\Xi^-$ & -0.98 & 1.01 \\
$\Lambda$ & -0.12 & 1.95 \\
$\Lambda\Sigma^0$ & 1.11 & 1.05
\end{tabular}
\end{table}

\section{WESS-ZUMINO-WITTEN SECTOR}

The Wess-Zumino-Witten (WZW) effective action accounts for the physics of the
chiral anomalies.\cite{WZW}  A lattice representation of the WZW action can be
constructed by converting derivatives to finite differences, just as was done
above for the non-anomalous meson and baryon sectors.
However, the WZW action is special in that its continuum form can be uniquely
derived --- with no unknown coefficients at leading chiral order --- from an
underlying fermion
action.  In fact, Aoki has shown that the unique continuum WZW action can be
derived from the {\em lattice} Wilson action.\cite{Ichinose,Aoki2}

Alternatively, one might be interested in defining a lattice effective action
that resides in the same discrete spacetime where the underlying fermion
action (Wilson, for example) resides.  However, a unique lattice WZW effective
action cannot be derived from the Wilson fermion action.
Here, we discuss this issue by closely following the work of
Refs.~\cite{Ichinose,Aoki2} but without invoking the continuum.

Begin with massless Wilson fermions coupled to external gauge fields.
\begin{eqnarray}
&\!\!& \hspace{-5mm}S^{(W)}(L,R) = \nonumber \\ &\!\!& \frac{1}{2}\sum_{n,\mu}(
\bar\psi^L_n\gamma_\mu L_{n,n+\mu}\psi^L_{n+\mu}
-\bar\psi^L_n\gamma_\mu L_{n,n-\mu}\psi^L_{n-\mu} \nonumber \\
&\!\!& +\bar\psi^R_n\gamma_\mu R_{n,n+\mu}\psi^R_{n+\mu}
-\bar\psi^R_n\gamma_\mu R_{n,n-\mu}\psi^R_{n-\mu}) \nonumber \\
&\!\!& -\frac{r}{2}\sum_{m,n}\left(\bar\psi^L_m\Box_{m,n}\psi^R_n
+\bar\psi^R_m\Box_{m,n}\psi^L_n\right).
\end{eqnarray}
Except for the chiral symmetry breaking Wilson term (proportional to $r$),
this action is invariant under local chiral transformations,
\begin{eqnarray}
\psi^L_n &\to& h_n\psi^L_n, \\
\psi^R_n &\to& g_n\psi^R_n, \\
L_{m,n}  &\to& h_mL_{m,n}h^\dagger_n, \\
R_{m,n}  &\to& g_mR_{m,n}g^\dagger_n,
\end{eqnarray}
where $h_n\in SU_L(N_f)$ and $g_n\in SU_R(N_f)$.
To obtain the WZW effective action, consider the axial transformation,
\begin{eqnarray}
\psi^L_n &\to& u_n^\dagger\psi^L_n, \\
\psi^R_n &\to& u_n\psi^R_n, \\
L_{m,n}  &\to& L_{m,n}^\prime\equiv u^\dagger_mL_{m,n}u_n, \\
R_{m,n}  &\to& R_{m,n}^\prime\equiv u_mR_{m,n}u^\dagger_n,
\end{eqnarray}
where $u_n^2 = U_n = \exp(-i\lambda^a\pi^a_n/F)$.
The WZW action is defined to be the difference
\begin{equation}
W_{WZW} = W(L^\prime,R^\prime) - W(L,R)
\end{equation}
where, as usual,
\begin{eqnarray}
Z(L,R) &=& \exp(-W(L,R)) \nonumber \\
       &=& \int{\cal D}\bar\psi{\cal D}\psi\,\exp\left(-S^{(W)}(L,R)\right).
\end{eqnarray}
After using $W=-\ln\det D=-{\rm Tr}\ln D$, inserting a complete set of states
into the trace, and performing some further algebra, one arrives at
\begin{eqnarray}
W_{WZW}&\!\!\!\!\!=\!\!\!\!\!&\frac{N_c}{4}\sum_n\int_0^1\!\!\!d\tau
       \int_{-\pi}^\pi\frac{d^4k}{(2\pi)^4}
          \sum_{m=0}^\infty\frac{{\rm Tr}_n(A_{m,n})}{(-s^2)^{m+1}}
       \nonumber \\ &\hspace{-2cm}& \label{Wwzw}
\end{eqnarray}
where $n$ sums over all lattice sites but $m$ is just an integer index.  Also,
\begin{eqnarray}
A_{m,n}&\!\!\!\!\!=\!\!\!\!\!&s\!\!/\!\left[\left(LT\!\!\!/P^L\!
      +\!RT\!\!\!/P^R\!+\!P^LQ_L\!+\!P^RQ_R\right)s\!\!/\right]^m \nonumber \\
      &\hspace{-2cm}& \times\frac{d~}{d\tau}\left[P^LQ_L+P^RQ_R\right]
\end{eqnarray}
and $\tau$ is the coordinate of Witten's extra dimension, so
$u^\tau_n\in SU(N_f)$ with $u^0_n=1$ and $u_n^1=u_n$.  We have also defined
\begin{eqnarray}
Q_L &=& -ru^\dagger\bigg(\sum_\mu Q^\mu\bigg)u^\dagger, \\
Q_R &=& -ru\bigg(\sum_\mu Q^\mu\bigg)u, \\
Q_{m,n}^\mu &=& \Box_{m,n}\cos(k_\mu) + i\partial_{m,n}\sin(k_\mu) \nonumber \\
             &&   - 2\delta_{m,n}(1-\cos(k_\mu)), \label{defQ} \\
T_{m,n}^\mu &=& 2\partial_{m,n}\cos(k_\mu) + i\Box_{m,n}\sin(k_\mu), \\
\partial_{m,n} &=& (\delta_{m+\mu,n}-\delta_{m-\mu,n})/2, \\
\Box_{m,n} &=& \delta_{m+\mu,n} + \delta_{m-\mu,n} - 2\delta_{m,n},
\end{eqnarray}
and $P_{R,L} = (1\pm\gamma_5)/2$.

In the continuum limit, $\partial_{m,n}$ and $\Box_{m,n}$ collapse to local
derivatives and the summations can be performed explicitly.\cite{Aoki2}
The expression for $W_{WZW}$ reproduces the well-known anomaly.
On a lattice, the summation over $m$ in Eq.~(\ref{Wwzw}) does not produce a
closed analytic form.  We have worked through the case of 2-dimensional
spacetime in some detail, and find that terms with arbitrarily many powers of
$Q^\mu$, as defined in Eq.~(\ref{defQ}), can contribute at low chiral orders.  The essential point is that
``$\lim_{a\to0}$'' and ``${\rm Tr}_n$'' do not commute.  
Notice that Tr$_n$ requires closed paths on the lattice.

\section{SUMMARY}

Lattice regularization can be applied to chiral perturbation theory.
The simplest lattice ChPT Lagrangian is obtained by converting derivatives
to finite differences in such a way as to preserve the desired symmetries
without introducing doublers or ghost states.  The Wess-Zumino-Witten sector of
the theory can be included in exactly the same way.
As a side issue, we have noted that
the known Wess-Zumino-Witten continuum coefficient,
$N_c/(240\pi^2)$, can be derived from the underlying lattice
fermion action\cite{Aoki2} but must then be retained by hand in lattice ChPT,
since a derivation that avoids the continuum step was found to be intractable.

As $a\to0$, ChPT observables are independent of the regularization scheme and
therefore the continuum limit of a lattice regularized calculation is identical
to the dimensional regularized result.
A determination of the differences between calculations at $a=0$ and
$a\neq0$ allows for a
discussion of the scheme dependence that does arise in practice, due to the
truncation of ChPT at some specific chiral order.  The lattice spacing is
easily adjusted in lattice ChPT and the $a$-dependences are thus obtained
directly.  Explicit verifications of scheme-independence, and conversely the
opportunity
to discover any unacceptably large scheme dependences that might exist in some
particular situation, add
confidence to the practical use of ChPT as an effective field theory.

\section*{ACKNOWLEDGEMENTS}

R.L. benefitted from discussions with Sinya Aoki, Maarten Golterman and
Derek Leinweber at the Workshop on Lattice Hadron Physics (Cairns, Australia)
where this work was presented.
P.O. is grateful to Wolfram Weise and the T-39 theory group at Technische
Universit\"at M\"unchen for their support and hospitality while a portion of
the research was in progress.
This work was also supported in part by the Deutsche Forschungsgemeinschaft
and the Natural Sciences and Engineering
Research Council of Canada.

\end{document}